# Separation of transport lifetimes in SrTiO$_3$-based two-dimensional electron liquids

Evgeny Mikheev[1,*], Christopher R. Freeze[1], Brandon J. Isaac[1], Tyler A. Cain[1], and Susanne Stemmer[1,*]

[1]Materials Department, University of California, Santa Barbara, CA 93106-5050, U.S.A.

*Correspondence to: emikheev@mrl.ucsb.edu or stemmer@mrl.ucsb.edu

**Abstract**

Deviations from Landau Fermi liquid behavior are ubiquitous features of the normal state of unconventional superconductors. Despite several decades of investigation, the underlying mechanisms of these properties are still not completely understood. In this work, we show that two-dimensional electron liquids at $SrTiO_3$/$R$$TiO_3$ ($R$ = Gd or Sm) interfaces reveal strikingly similar physics. Analysis of Hall and resistivity data show a clear separation of transport and Hall scattering rates, also known as "two-lifetime" behavior. This framework gives a remarkably simple and general description of the temperature dependence of the Hall coefficient. Distinct transport lifetimes accurately describe the transport phenomena irrespective of the nature of incipient magnetic ordering, the degree of disorder, confinement, or the emergence of non-Fermi liquid behavior. The Hall scattering rate diverges at a critical quantum well thickness, coinciding with a quantum phase transition. Collectively, these results introduce new constraints on the existing microscopic theories of lifetime separation and point to the need for unified understanding.




# I. Introduction

Understanding the anomalous normal-state transport properties of unconventional superconductors remains one of the most challenging problems in condensed matter physics [1-3]. Prominent manifestations include strong deviations from conventional metallic Fermi liquid behavior in the temperature ($T$) dependence of the electrical resistivity ($\rho_{xx}$), combined with a non-trivial temperature dependence of the Hall coefficient. In 1991, Ong, Anderson, and collaborators suggested that a unified description could be obtained if the longitudinal conductivity ($\sigma_{xx}$) and the Hall conductivity ($\sigma_{xy}$) contain two distinct scattering rates, $\tau_{tr}$ and $\tau_H$, so that $\sigma_{xx} \sim \tau_{tr}$ and $\sigma_{xy} \sim \tau_{tr}\tau_H$ [4-7]. This description represents a radical departure from Boltzmann transport theory of a normal, isotropic metal, where the longitudinal and Hall conductivities are determined by a single quasiparticle scattering rate, $\tau$, so that $\sigma_{xx} \sim \tau$ and $\sigma_{xy} \sim \tau^2$, resulting in a $\tau$-independent Hall coefficient $R_H = \sigma_{xy}/\sigma_{xx}^2$. The most intriguing experimental signature suggesting a lifetime separation is the temperature dependence of the Hall angle of the cuprate superconductors, $\cot(\theta_H) = \sigma_{xx}/\sigma_{xy} \sim \tau_H$, which follows Fermi-liquid behavior $\left[\cot(\theta_H) = C + \alpha T^2\right]$. This is in sharp contrast to the non-Fermi liquid behavior of $\rho_{xx}$ ($\rho_{xx} = \rho_0 + AT^n$ with $n < 2$). Supporting evidence for an independent scattering rate $\tau_H$ comes from impurity effects in the cuprates [4, 8-10]: introduction of disorder into the CuO$_2$ planes increases $C$, while $\alpha$ is barely affected. Two lifetimes also describe several heavy fermion systems [11-13]. Furthermore, V$_2$O$_{3-x}$, a three-dimensional antiferromagnet, was also reported to have a well-defined $T^2$ dependence of $\cot(\theta_H)$, while exhibiting a non-Fermi liquid $\rho_{xx}$ with $n = 3/2$ [14].



Despite its success in describing the observed behavior, the microscopic physics of two distinct lifetimes is extremely challenging [6, 7]. The original proposal by Anderson [5] involved decoupling of electron spin and charge into spinon and holon quasiparticles in a two-dimensional Luttinger liquid. More conservative models have also been put forward that do not require the existence of two separate scattering rates, $\tau_{tr}$ and $\tau_H$. These are based on scattering anisotropies on the Fermi surface [15-17], as, for example, caused by antiferromagnetic spin fluctuations [18-20]. Despite concerns about the lack of generality and the necessary reliance on very specific cancellations in the scattering rates [7], such scenarios can describe experimental data from the cuprates [21].

In this work, we demonstrate clear manifestations of two distinct lifetimes in a very different system: the two-dimensional electron liquid (2DEL) at SrTiO$_3$/$R$TiO$_3$ interfaces ($R$ = Gd or Sm). Systematic tuning of the boundary conditions establishes that the lifetime separation is pervasive: it is observed independent of the type of (incipient) magnetic order in the 2DEL, its presence is insensitive to the degree of disorder and confinement, and it describes both non-Fermi liquids (NFL) and Fermi liquids (FL). The ubiquity of scattering rate separation introduces constraints on the underlying origins and emphasizes the need for a more unified microscopic theory. We furthermore show that the strongest manifestation of the two-lifetime behavior occurs near a quantum critical point (QCP) in this system.

## II. Experimental

$R$TiO$_3$/SrTiO$_3$/$R$TiO$_3$ quantum well structures were grown on (001) (La$_{0.3}$Sr$_{0.7}$)(Al$_{0.65}$Ta$_{0.35}$)O$_3$ (LSAT) substrates by hybrid molecular beam epitaxy [22, 23]. Each SrTiO$_3$/$R$TiO$_3$ interface in



the $R$TiO$_3$/SrTiO$_3$/$R$TiO$_3$ quantum well structures electrostatically introduces ~3.4×10$^{14}$ cm$^{-2}$ carriers into the SrTiO$_3$, which compensate for the interfacial polar discontinuity (Fig. S1) [24, 25]. The spatial confinement of the ~6.8×10$^{14}$ cm$^{-2}$ carriers, the effective three-dimensional carrier density, and the degree to which the two 2DELs overlap, is determined by the SrTiO$_3$ thickness ($t_{QW}$), which can be controlled with single-layer precision [26, 27]. We specify $t_{QW}$ by the number of SrO layers in the quantum well. Quantum wells in *ferrimagnetic* GdTiO$_3$ (Curie temperature ~ 30 K) show ferromagnetism below a critical $t_{QW}$ of 5 SrO layers, followed by a transition to a correlated insulator at 2 SrO layers that is accompanied by a symmetry-lowering structural transition [26-29]. Quantum wells in *antiferromagnetic* SmTiO$_3$ (Neel temperature ~ 50 K) remain metallic down to a single SrO layer and exhibit non-Fermi-liquid (NFL) behavior [30]. Thus $t_{QW}$ tunes the proximity of the electron system to magnetic order and, as shown below, across a QCP. SrTiO$_3$ layer thicknesses (in number of SrO layers) were calibrated using scanning transmission electron microscopy [27, 29]. Barrier thicknesses were nominally 10 nm (for SmTiO$_3$) and 4 nm (for GdTiO$_3$) on either side of the SrTiO$_3$ quantum well. Electrical contacts consisted of 40 nm Ti/400 nm Au, deposited by electron beam evaporation using shadow masks, in either Van der Pauw geometry or Hall bar geometry. Resistances as a function of temperature and magnetic field were measured using Physical Property Measurement Systems (Quantum Design PPMS Dynacool and PPMS).

## II. Results

Figure 1 shows the longitudinal sheet resistance $R_{xx}$, the apparent carrier density $(eR_H)^{-1}$, and $\cot(\theta_H)$, as a function of temperature and $t_{QW}$ for SmTiO$_3$/SrTiO$_3$/SmTiO$_3$ (top row) and



GdTiO$_3$/SrTiO$_3$/GdTiO$_3$ (bottom row) samples. The temperature dependence of $R_{xx}$ can be described as follows:

$$R_{xx} = R_0 + AT^n = \frac{1}{eN\mu_{tr}(T)}, \qquad (1)$$

where $R_0$ is the residual resistance and $A$ the temperature coefficient. For quantum wells in SmTiO$_3$, the exponent $n$ deviates from the FL value ($n = 2$) [25, 30] and is as low as $n = 1.6$ in thin quantum wells, see Fig. 2(d) and Fig. S2. The second equality is the Drude model, where $N$ is the sheet carrier density, and the longitudinal transport mobility is $\mu_{tr} = e\tau_{tr}/m^*$ ($m^*$ is the effective electron mass).

$(eR_H)^{-1}$ depends in a non-trivial fashion on temperature and $t_{QW}$. It converges towards the actual carrier density of $\sim 7 \times 10^{14}$ cm$^{-2}$ only at high temperatures. Furthermore, we note the discontinuity with decreasing $t_{QW}$ for both types of quantum wells: $(eR_H)^{-1}$ first increases down to 5 SrO layers and then decreases, and this behavior is particularly pronounced at low temperatures.

The Hall effect is linear in magnetic field ($H$) up to 9 T at all measured temperatures. The absence of anomalous contributions allows us to use the field-independent Hall angle, $H\cot(\theta_H) = H R_{xx}/R_{xy} = R_{xx}/R_H$, which corresponds to the inverse Hall carrier mobility ($\mu_H^{-1}$). As shown in Figs. 1(e,f), $H\cot(\theta_H)$ follows a well-defined $T^2$ dependence from $\sim 50$ K up to room temperature *for all samples*, including those that show NFL behavior in $R_{xx}$:



$$H \cot(\theta_H) = \frac{R_{xx}}{R_H} = \mu_H^{-1} = H(C + \alpha T^2). \tag{2}$$

Combining Eqs. (1) and (2) describes the temperature-dependence of $(eR_H)^{-1}$:

$$\frac{1}{eR_H} = N \frac{\mu_{tr}}{\mu_H} = \frac{H}{e} \frac{C + \alpha T^2}{R_0 + AT^n}. \tag{3}$$

Eq. (3) is the two-lifetime description: it rationalizes the temperature dependence of $R_H$ as the ratio between the two scattering rates, $\tau_{tr}$ and $\tau_H$.

The fitting procedure for resistance and Hall data shown in Fig. 1 involved a two-step process. First, the linear portion in the $R_{xx}/R_H$ vs. $T^2$ plot is fit to Eq. (2), using $C$ and $\alpha$ as adjustable parameters. The $R_H$ vs. $T$ data is then fitted to Eq. (3), with $R_0$, $A$ and $n$ as adjustable parameters, while $C$ and $\alpha$ are fixed at values obtained in the first step. For $R$ = Gd, the fit of $R_H^{-1}$ was carried out over the entire measured temperature range (2 – 300 K). For $R$ = Sm and thin quantum wells (< 5 SrO layers), data below 50 K had to be excluded from the fit, as Eq. (3) does not account for the physics that causes the downturn in $(eR_H)^{-1}$ at low temperatures for these samples. We attribute to the downturn to a true reduction in mobile charge carrier density $N$. This is only observed for thin quantum wells embedded in antiferromagnetic SmTiO$_3$. A partial gap opening may be consistent with itinerant antiferromagnetism (spin density wave) in thin quantum wells in SmTiO$_3$, which would also be consistent with the pseudogaps observed in thin quantum wells [31]. Furthermore, weak negative magnetoresistance that is not consistent with weak localization points to magnetic fluctuations [30].



Self-consistency of the fit was checked by comparing the temperature dependence of $R_{xx}$ calculated using the fit parameters and $R_{xx} = R_0 + AT^n$, with the measured $R_{xx}$. In other words, all data in Fig. 1 was fitted with a single set of adjustable parameters for each sample.

As is evident from fits shown as solid lines in Figs. 1(c,d), Eq. (3) provides an extremely good description of $(eR_H)^{-1}$ as a function of temperature and $t_{QW}$, for both types of quantum wells, up to 300 K. It describes the diverging low-temperature $(eR_H)^{-1}$ for both types of quantum wells and the discontinuity with $t_{QW}$. The fit parameters in Eq. (3) are the residual resistances, $R_0$ and $C$, the temperature coefficients $A$ and $\alpha$, and $n$. It is important to note that $R_0$, $A$ and $n$ extracted by fitting $R_{xx}/R_H$ and $R_H$ [Eqs. (2) and (3)] agree well with fits using the $R_{xx}$ data alone [solid lines in Figs. 1(a,b)]. The upturn in $R_{xx}$ seen at low temperature can be attributed to the loss of charge carriers, discussed above, and was not included in the fits.

In the presence of NFL behavior, the high temperature limit of the Hall coefficient is temperature dependent: $(eR_H)^{-1} \sim T^{2-n}$. This is seen for thin SmTiO$_3$ quantum wells, which have a small positive slope near room temperature. More generally, the interplay between the two scattering rates can give $\mu_{tr}/\mu_H \neq 1$ in Eq. (3) and result in under- or overestimation of carrier density from the Hall effect.

In the $T = 0$ K limit, $(eR_H)^{-1}$ is determined by $C/R_0$, i.e., the ratio of the residuals in $\tau_H$ and $\tau_{tr}$:

$$\frac{1}{eR_H(0\ \text{K})} = N\frac{\mu_{tr}(0\ \text{K})}{\mu_H(0\ \text{K})} = \frac{H}{e}\frac{C}{R_0}. \tag{4}$$



$C/R_0$ extracted from fitting the temperature dependence matches very well with the experimental $\left(eR_H\right)^{-1}$ at low temperature, as shown in Fig. 2(a). An intriguing observation is the diverging $C/R_0$ at $t_{QW} \sim 5$ SrO layers, which is the origin of the discontinuity in $\left(eR_H\right)^{-1}$ mentioned above. A diverging $T = 0$ K Hall effect suggests a possible quantum critical point (QCP) and/or a Lifshitz transition at this thickness. Furthermore, the divergence in $C/R_0$ coincides with the appearance of other phenomena [Figs. 2(d,f)], specifically:

- Discontinuity in $n$: quantum wells in antiferromagnetic SmTiO$_3$ transition to NFL behavior in $\tau_{tr}$ and $R_{xx}$ (but not the Hall angle) below 5 SrO layers [Fig. 2(d)], with $n \sim 1.6$ for 4 SrO layers. In contrast, $n$ remains close to $\sim 2$ for quantum wells in GdTiO$_3$ at all thicknesses {Fig. 2(d) and ref. [26]}.

- Onset of magnetic order: quantum wells in GdTiO$_3$ become ferromagnetic [28, 32]. The Curie temperature is $\sim 10$ K for 3 SrO layers, decreases with increasing $t_{QW}$, and ferromagnetism is not detectable for 6 or more SrO layers [32]. The downturn in $\left(eR_H\right)^{-1}$ that appears at low temperature in quantum wells in SmTiO$_3$ at $t_{QW} \sim 5$ SrO layers is consistent with a loss of carriers, for example due to a spin density wave gap opening [see Fig. 1(c)].

To further clarify the origin of the divergence in $C/R_0$, Figs. 2(b) and (c) show $R_0$ and $C$ as a function of $t_{QW}$. Both $R_0$ and $C$ scale with disorder [4, 8]. In quantum wells, interface roughness scattering is a dominant disorder contribution [33], so both $R_0$ and $C$ increase with decreasing $t_{QW}$. $R_0$ increases more quickly for quantum wells in GdTiO$_3$: this likely reflects the larger octahedral distortions and closer proximity to the metal-insulator transition [27]. This



affects the Fermi surface areas and thereby $R_0$ [34, 35]. Such monotonic contributions dominate the thickness dependence of $R_0$. In the case of $C$, there is an additional contribution, which is divergent near $t_{QW} \sim 5$ SrO. This is particularly visible when multiplying $C$ and $R_0$ by $t_{QW}$ as shown in the insets in Figs. 2(b) and (c), respectively. This removes the contribution of the approximately linear increase with decreasing $t_{QW}$ [the increase in both quantities at large $t_{QW}$ is caused by the fact that the actual extent of the two 2DELs is less than the physical quantum well thickness, $t_{QW}$]. The divergence in $C$ occurs at the same 5 SrO layer thickness for both types of quantum wells, $R$ = Gd and Sm.

The temperature coefficient $\alpha$ of $\tau_H$ is continuous across the entire thickness range [25]. The corresponding coefficient for longitudinal transport ($A$) has an apparent jump near the QCP (Fig. S3) [25], but it is due to the change in $n$. Thus, the two quantities that are discontinuous/divergent near $t_{QW} \sim 5$ SrO layers are $n$ (for quantum wells in SmTiO$_3$) and $C$ (for both).

## IV. Discussion

The lifetime separation provides a remarkably complete description of the transport properties of SrTiO$_3$ quantum wells. Similar to the cuprates, we find that the Hall angle follows a $T^2$ temperature dependence, even when $R_{xx}$ does not. The clear separation of $C$ and $R_0$ adds to the existing evidence that $\tau_{tr}$ and $\tau_H$ are indeed distinct scattering rates in these systems, with different underlying physics that influences them. We discuss the implications of the results with regards to proposed explanations of the two-lifetime behavior and the phase behavior in this system.

### A. Scattering rate anisotropy



For the cuprates, it has been suggested that the transport scattering rates can *appear separated* due to anisotropy in the scattering rates on the Fermi surface [18-20], i.e., sections with high ("hot spots") and low ("cold spots") scattering rates. This leads to a breakdown of the results obtained with an isotropic scattering rate [36]. The simplest case is a Fermi surface where the isotropic scattering rate ($\tau_1$) is locally increased (hot spot) or decreased (cold spot) to yield a scattering rate $\tau_2$ [15, 16]. $\sigma_{xx}$ is proportional to the integrated average along the Fermi surface contour; a narrow hot or cold spot is negligible, thus $\sigma_{xx} \sim \tau_1$. $\sigma_{xy}$ is approximately proportional to the product of the two scattering rates: $\sigma_{xy} \sim \tau_1 \tau_2$ [15]. This gives $\cot(\theta_H) \sim \tau_2$, leading to a scattering rate separation of form $\tau_{tr} = \tau_1$ and $\tau_H = \tau_2$. In the cuprates the scattering anisotropy [21, 37] is broadened (cosine-like), which causes a mixing of $\tau_1$ and $\tau_2$ in the transport integrals. Approximate dependencies $\sigma_{xx} \sim \sqrt{\tau_1 \tau_2}$, $\sigma_{xy} \sim \tau_1 \sqrt{\tau_1 \tau_2}$ and $\cot(\theta_H) \sim 1/\tau_1$ were found by integration over realistic Fermi surfaces [19]. This recovers the correct temperature behavior for the cuprate normal state, where $\sigma_{xx} \sim T^{-1}$ and $\cot(\theta_H) \sim T^2$, with Landau Fermi liquid behavior across most of the Fermi surface $(\tau_1 \sim T^{-2})$.

SrTiO$_3$-based 2DELs contain highly anisotropic Fermi surfaces associated with $d_{xz,yz}$-derived bands [38-40]. Given the clear separation of $C$ and $R_0$, narrow, clearly separated scattering rates would be required *for all samples* (i.e., no mixing of $\tau_1$ and $\tau_2$) to describe the data. In the cuprates, antiferromagnetic spin fluctuations can result in anomalously high scattering at certain points in the Brillouin zone. Here, two lifetimes are found independent of the particular magnetic order parameter. Thin SrTiO$_3$ 2DELs in GdTiO$_3$ are ferromagnets with a $T_c \sim 10$ K. The wavevectors of ferromagnetic fluctuations should be large, and scattering should affect the entire Fermi surface. This is in sharp contrast to the incommensurate wave vectors



typical for antiferromagnetic fluctuations considered for the cuprates, which would create the required scattering rate anisotropy. Thus, *if* strong scattering rate anisotropies, having different temperature dependencies, are the cause of lifetime separation in this system, they are likely not related to the spin fluctuations. Unlike the lifetime separation, the specific type of magnetic order parameter *does* determine the magnitude of temperature exponent *n*, which changes to a NFL exponent at the QCP only for the (nearly) antiferromagnetic quantum wells.

### B. Connection to a quantum critical point

The ratio of the residuals, $C/R_0$, corresponds to the 0 K limit of $(eR_H)^{-1}$ [Eq. (4)]. It diverges at a critical $t_{QW}$ in this system. This has intriguing implications for a quantum critical point in this system, and possibly the nature of the lifetime separation itself. In general, the temperature dependence of $(eR_H)^{-1}$ may be taken as a transition from a high temperature regime, where strong electron-electron scattering dominates [$(eR_H)^{-1} \sim \alpha/A$ in Eq. (3)], to low temperature, where $(eR_H)^{-1}$ is dominated by $C/R_0$. The 0 K limit of $(eR_H)^{-1}$ originates from disorder scattering, which is sensitive to changes in the electronic structure (Fermi surface sizes and topology). As discussed above, this explains the monotonous increase in both *C* and $R_0$ with decreasing $t_{QW}$. The *divergence of $C/R_0$*, however, is caused by the residual *C* of the Hall angle (inverse of the Hall mobility). While $R_0$ is sensitive to the Fermi surface topology [34], *C* is the residual of the inverse of the Hall mobility and thus only sensitive to the quasiparticle mass and scattering length. This indicates that the divergence of *C/R_0 is not caused by an abrupt* change in band structure (Fermi surface) at 5 SrO thickness, as this should be reflected in $R_0$ [34]. Thus the



QCP in this system does not appear to be a Lifshitz transition. Rather, the divergence of $C$ (the 0 K limit of $\tau_H$) points to a diverging mass or 0-K-scattering length at the critical $t_{QW}$ at which magnetic order and NFL behavior appear. It is possible that the critical $t_{QW}$ corresponds to the thickness at which the two 2DELs at each interface begin to strongly overlap. Similar physics would then lie at the origins of the two-lifetime behavior in this system, which becomes strongest (most pronounced separation of $C$ and $R_0$) near the QCP. A recent theory suggests that the DC conductivity is dominated by the critical point, while the Hall angle is dominated by the Umklapp scattering and shows Fermi liquid behavior [41]. While the residuals ($C$ and $R_0$) were not considered in this theory, it does establish a connection between the proximity to a QCP and lifetime separation, which appears to be qualitatively in agreement with the findings reported here. In this context, it would also be interesting to carry out a similar analysis of the Hall data on other thin film systems were proximity to quantum critical point has been suggested [42, 43].

It is important to note that the results strongly support the notion that the non-trivial thickness and temperature dependences of $(eR_H)^{-1}$ are not caused by (changes in) the band structure *per se*. Specifically, the temperature dependence of $(eR_H)^{-1}$ reflects the transition from electron-electron scattering that dominates the mobility at high temperatures to one at low temperatures that it is controlled by another type of interaction/scattering that underlies the diverging $C/R_0$. The relative insensitivity of the $(eR_H)^{-1}$ to details of the band structure explains the "single band" picture of $(eR_H)^{-1}$ that describes the data even though transport occurs in multiple subbands in this system [44, 45]. We note that in any case, multiband models would have enormous difficulties in describing the temperature dependence of $(eR_H)^{-1}$. For example,



transport in multiple bands containing only electrons can only result in a value of $(eR_H)^{-1}$ that is *lower* (not *higher*) than the true carrier density *N*, as can be shown by a simple analysis of the low field multicarrier expression, $R_H = -\sum_i n_i \mu_i^2 / e \left(\sum_i n_i \mu_i\right)^2$, where the subscript indicates the subband index. Here, low temperature values of $(eR_H)^{-1}$ exceed the true carrier density for certain $t_{QW}$. We note that the assumption of only electron-like, $d_{xy}$ and $d_{xz,yz}$-derived bands in this system is supported by angle-resolved x-ray photoemission spectrosopy (ARPES) of similar interfaces [46], recent DFT studies of quantum wells in this and related systems [45, 47], as well as by preliminary ARPES data on the quantum wells in this study [48].

A real change in Fermi surface *as a function of temperature* in thin quantum wells is, however, a likely explanation for the only feature of $R_H(T)$ that is not captured by the two-life model: the downturn of $(eR_H)^{-1}$ below 50 K in thin quantum wells in $SmTiO_3$ Fig. 1(c). A density wave gap could explain the loss of carriers at low temperatures.

## V. Conclusions

In summary, the results emphasize the need for a microscopic theory of the lifetime separation that is applicable to a wider range of systems than previously considered. Analysis of the conditions leading to a pronounced lifetime separation – a quantum critical point that separates the 0 K residuals in the longitudinal resistance and Hall angle – suggests that such a microscopic theory would have wide-ranging implications for the origins of anomalous transport behavior. The results add strong support to the notion that "strange metal behavior", such as the



appearance of two distinct lifetimes, emerges out of a Fermi liquid in which electron-electron scattering is strong – this is *the* common feature of all electron systems that exhibit lifetime separation. Artificially engineered structures, such as the $SrTiO_3/RTiO_3$ system studied here, are a fascinating playground for studying the normal state physics present in unconventional superconductors. We hope that this work will be stimulating input for the broader debate on the origin of anomalous transport phenomena in strongly correlated materials.


**Acknowledgements**

We gratefully acknowledge many helpful discussions with Alessandra Lanzara, Piers Coleman, Andrew Schofield, Jim Allen, and Leon Balents. We also thank Santosh Raghavan and Clayton Jackson for the growth of some of the samples that were analyzed in this study. We acknowledge support from the U.S. Army Research Office (grant nos. W911-NF-09-1-0398 and W911NF-14-1-0379). Acquisition of the oxide MBE system used in this study was made possible through an NSF MRI grant (Award No. DMR 1126455). The work also made use of the UCSB Nanofabrication Facility, a part of the NSF-funded NNIN network and of Central facilities supported by the MRSEC Program of the U.S. National Science Foundation under Award No. DMR 1121053.





# References

[1]  P. W. Anderson, Proc. Natl. Acad. Sci. **92**, 6668 (1995).

[2]  A. J. Schofield, Contemp. Phys. **40**, 95 (1999).

[3]  P. A. Lee, N. Nagaosa, and X. G. Wen, Rev. Mod. Phys. **78**, 17 (2006).

[4]  T. R. Chien, Z. Z. Wang, and N. P. Ong, Phys. Rev. Lett. **67**, 2088 (1991).

[5]  P. W. Anderson, Phys. Rev. Lett. **67**, 2092 (1991).

[6]  A. T. Zheleznyak, V. M. Yakovenko, H. D. Drew, and I. I. Mazin, Phys. Rev. B **57**, 3089 (1998).

[7]  P. Coleman, A. J. Schofield, and A. M. Tsvelik, J Phys-Condens Mat **8**, 9985 (1996).

[8]  X. Gang, X. Peng, and M. Z. Cieplak, Phys. Rev. B **46**, 8687 (1992).

[9]  Y. S. Wu, C. R. Becker, A. Waag, R. N. Bicknelltassius, and G. Landwehr, Semicond Sci Tech **8**, S293 (1993).

[10] J. M. Harris, Y. F. Yan, and N. P. Ong, Phys. Rev. B **46**, 14293 (1992).

[11] S. Nair, S. Wirth, S. Friedemann, F. Steglich, Q. Si, and A. J. Schofield, Adv Phys **61**, 583 (2012).

[12] Y. Nakajima, K. Izawa, Y. Matsuda, S. Uji, T. Terashima, H. Shishido, R. Settai, Y. Onuki, and H. Kontani, J Phys Soc Jpn **73**, 5 (2004).

[13] Y. Nakajima, H. Shishido, H. Nakai, T. Shibauchi, K. Behnia, K. Izawa, M. Hedo, Y. Uwatoko, T. Matsumoto, R. Settai, Y. Onuki, H. Kontani, and Y. Matsuda, J Phys Soc Jpn **76**, 024703 (2007).

[14] T. F. Rosenbaum, A. Husmann, S. A. Carter, and J. M. Honig, Phys. Rev. B **57**, R13997 (1998).

[15] A. Carrington, A. P. Mackenzie, C. T. Lin, and J. R. Cooper, Phys. Rev. Lett. **69**, 2855 (1992).





[16]  C. Kendziora, D. Mandrus, L. Mihaly, and L. Forro, Phys. Rev. B **46**, 14297 (1992).

[17]  N. P. Ong, Phys. Rev. B **43**, 193 (1991).

[18]  R. Hlubina, and T. M. Rice, Phys. Rev. B **51**, 9253 (1995).

[19]  B. P. Stojkovic, and D. Pines, Phys. Rev. B **55**, 8576 (1997).

[20]  L. B. Ioffe, and A. J. Millis, Phys. Rev. B **58**, 11631 (1998).

[21]  N. E. Hussey, J Phys-Condens Mat **20**, 123201 (2008).

[22]  B. Jalan, R. Engel-Herbert, N. J. Wright, and S. Stemmer, J. Vac. Sci. Technol. A **27**, 461 (2009).

[23]  P. Moetakef, J. Y. Zhang, S. Raghavan, A. P. Kajdos, and S. Stemmer, J. Vac. Sci. Technol. A **31**, 041503 (2013).

[24]  P. Moetakef, T. A. Cain, D. G. Ouellette, J. Y. Zhang, D. O. Klenov, A. Janotti, C. G. Van de Walle, S. Rajan, S. J. Allen, and S. Stemmer, Appl. Phys. Lett. **99** (2011).

[25]  See Supplementary Information at [link to be inserted by publisher] for a schematic of the quantum well structure, analysis of the non-Fermi liquid behavior of $R_{xx}$, and additional results for the fit parameters as a function of quantum well thickness.

[26]  P. Moetakef, C. A. Jackson, J. Hwang, L. Balents, S. J. Allen, and S. Stemmer, Phys. Rev. B **86**, 201102 (2012).

[27]  J. Y. Zhang, C. A. Jackson, R. Chen, S. Raghavan, P. Moetakef, L. Balents, and S. Stemmer, Phys. Rev. B **89** (2014).

[28]  C. A. Jackson, and S. Stemmer, Phys. Rev. B **88**, 180403 (2013).

[29]  J. Y. Zhang, J. Hwang, S. Raghavan, and S. Stemmer, Phys. Rev. Lett. **110**, 256401 (2013).

[30]  C. A. Jackson, J. Y. Zhang, C. R. Freeze, and S. Stemmer, Nat. Commun. **5**, 4258 (2014).





[31] P. B. Marshall, B. J. Isaac, E. Mikheev, S. Raghavan, and S. Stemmer, Manuscript in preparation. (2015).

[32] P. Moetakef, J. R. Williams, D. G. Ouellette, A. P. Kajdos, D. Goldhaber-Gordon, S. J. Allen, and S. Stemmer, Phys Rev X **2**, 021014 (2012).

[33] A. Gold, Phys. Rev. B **35**, 723 (1987).

[34] T. M. Rice, and W. F. Brinkman, Phys. Rev. B **5**, 4350 (1972).

[35] M. Imada, A. Fujimori, and Y. Tokura, Rev. Mod. Phys. **70**, 1039 (1998).

[36] J. S. Dugdale, and L. D. Firth, Phys. kondens. Materie **9**, 54 (1969).

[37] M. Abdel-Jawad, M. P. Kennett, L. Balicas, A. Carrington, A. P. Mackenzie, R. H. McKenzie, and N. E. Hussey, Nat. Phys. **2**, 821 (2006).

[38] S. Y. Park, and A. J. Millis, Phys. Rev. B **87**, 205145 (2013).

[39] G. Khalsa, and A. H. MacDonald, Phys. Rev. B **86**, 125121 (2012).

[40] S. J. Allen, B. Jalan, S. Lee, D. G. Ouellette, G. Khalsa, J. Jaroszynski, S. Stemmer, and A. H. MacDonald, Phys. Rev. B. **88**, 045114 (2013).

[41] M. Blake, and A. Donos, Phys. Rev. Lett. **114**, 021601 (2014).

[42] J. Liu, M. Kargarian, M. Kareev, B. Gray, P. J. Ryan, A. Cruz, N. Tahir, Y. D. Chuang, J. H. Guo, J. M. Rondinelli, J. W. Freeland, G. A. Fiete, and J. Chakhalian, Nat. Commun. **4**, 2714 (2013).

[43] S. J. Allen, A. J. Hauser, E. Mikheev, J. Y. Zhang, N. E. Moreno, J. Son, D. G. Ouellette, J. Kally, A. Kozhanov, L. Balents, and S. Stemmer, Accepted for publication in APL Mater. (2015).

[44] F. Lechermann, and M. Obermeyer, arXiv:1411.1637 [cond-mat.mtrl-sci] (2014).





[45]  L. Bjaalie, A. Janotti, B. Himmetoglu, and C. G. Van de Walle, Phys. Rev. B **90**, 195117 (2014).

[46]  Y. J. Chang, L. Moreschini, A. Bostwick, G. A. Gaines, Y. S. Kim, A. L. Walter, B. Freelon, A. Tebano, K. Horn, and E. Rotenberg, Phys. Rev. Lett. **111**, 126401 (2013).

[47]  D. Doennig, and R. Pentcheva, Sci. Rep. **5**, 79090 (2015).

[48]  A. Lanzara, private communication.




**Figure Captions**

**Figure 1:** Transport data for $R$TiO$_3$/SrTiO$_3$/$R$TiO$_3$ quantum wells as a function of temperature with $R$ = Sm (top row) and Gd (bottom row). (a,b) Longitudinal resistance $R_{xx}$, (c,d) inverse of the Hall coefficient, $(eR_H)^{-1}$ and (e,f) Hall angle $R_{xx}/R_H$ on a $T^2$ scale. The labels indicate the SrTiO$_3$ quantum well thickness in terms of the number of SrO layers they contain. The solid lines are fits to the data, using Eqs. (1-3) and a single set of adjustable parameters.

**Figure 2:** Transport parameters as a function of SrTiO$_3$ thickness for $R$ = Sm (blue) and Gd (red). (a) $(eR_H)^{-1}$ in the $T$ = 0 K limit, from fitting to Eq. (3) (full symbols) and measured $(eR_H)^{-1}$ at $T$ = 2 K. (b,c) Residuals $R_0$ and $C$. The insets show the same quantities multiplied by the $t_{QW}$. (d) Temperature exponent $n$ in $R_{xx}$. (e,f) Transitions between FL, NFL and insulator (top row) and magnetic ordering in the quantum well (bottom row).



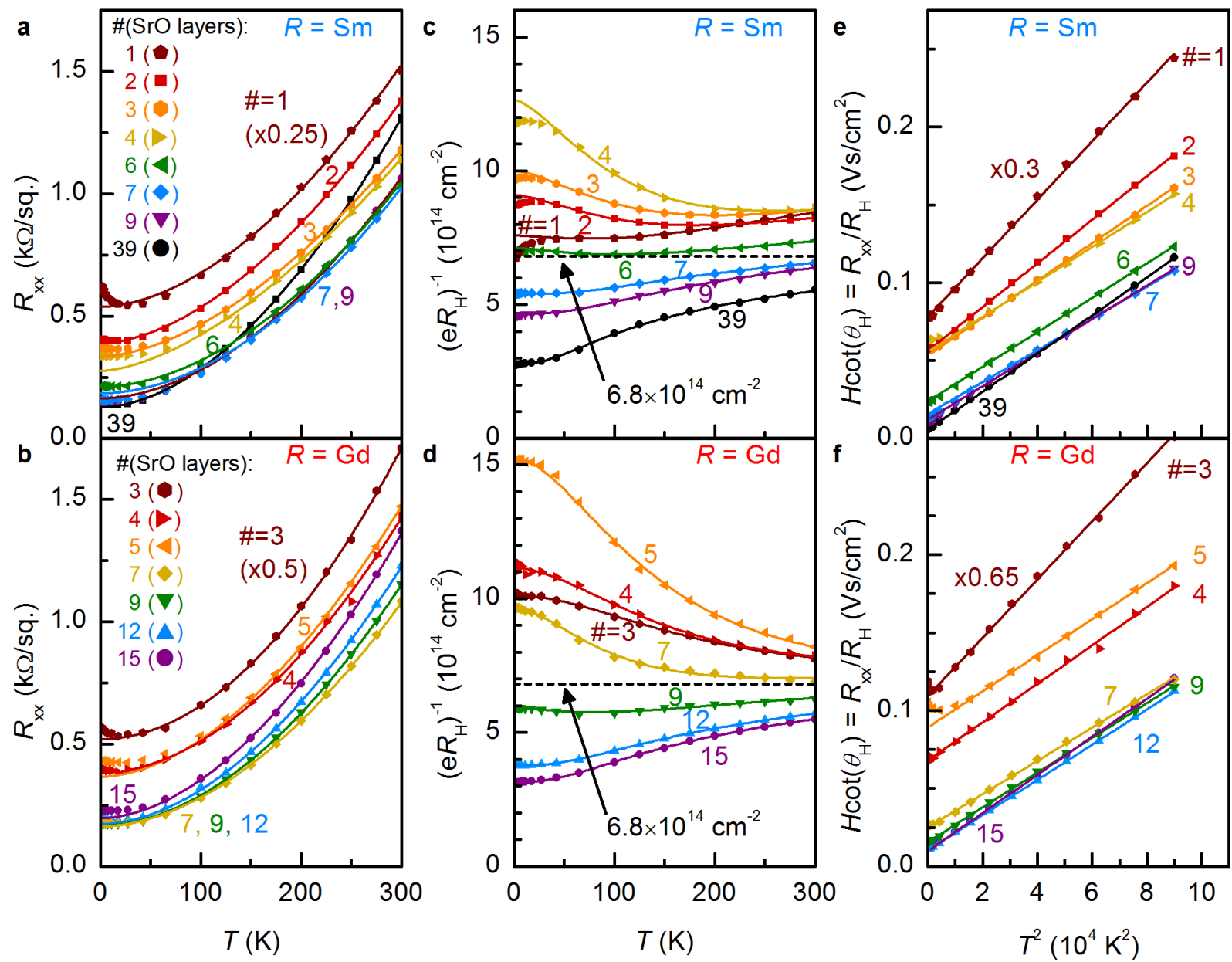

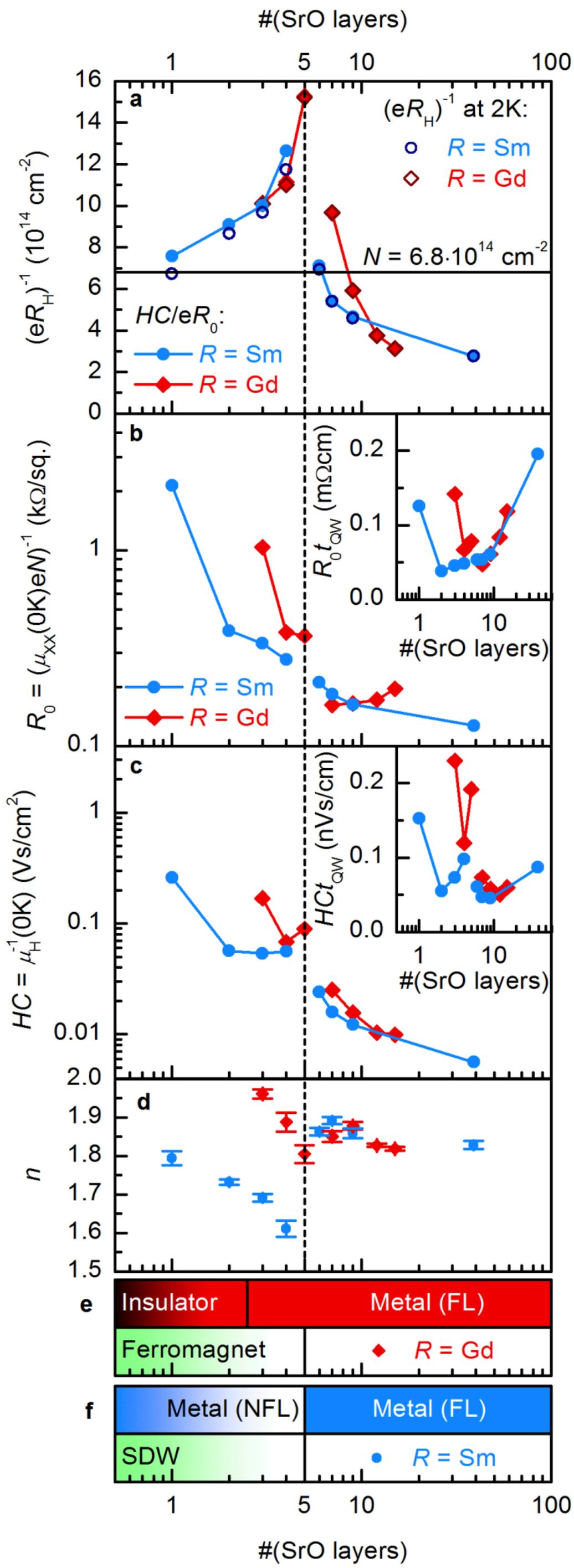

## Supplementary Information

# Separation of transport lifetimes in SrTiO$_3$-based two-dimensional electron liquids


Evgeny Mikheev, Christopher R. Freeze, Brandon J. Isaac, Tyler A. Cain, and Susanne Stemmer

Materials Department, University of California, Santa Barbara, CA 93106-5050, U.S.A.


Figure S1 shows a schematic of the $R$TiO$_3$/SrTiO$_3$/$R$TiO$_3$ quantum wells and the mechanism of 2DEL formation. $R$TiO$_3$/SrTiO$_3$ is a polar/nonpolar interface: SrO and TiO$_2$ layers within SrTiO$_3$ are charge neutral, but the $R$O and TiO$_2$ layers in $R$TiO$_3$ carry formal +1 and -1 charges. The TiO$_2$ layer at the interface receives ½ of an electron per cubic surface unit cell from the terminating $R$O layer, which forms the mobile 2DEL with a sheet carrier density of ~ $3.4 \times 10^{14}$ cm$^{-2}$. The mobile charge mostly confined within a few nanometers from the interface. In $R$TiO$_3$/SrTiO$_3$/$R$TiO$_3$ structures, there are two 2DELs (top and bottom) giving a total charge density of ~$6.8 \times 10^{14}$ cm$^{-2}$. The degree to which these 2DELs overlap and the three-dimensional carrier density is controlled by the SrTiO$_3$ thickness.

Figure S2 shows the temperature dependence of $R_{xx}$ and the degree to which it deviates from the $T^2$ behavior of a Landau Fermi liquid (FL):

$$R_{xx} = R_0 + AT^n. \tag{S1}$$

For GdTiO$_3$-based quantum wells, the exponent $n$ is close to 2, as illustrated by the linearity of $R_{xx}$ when plotted against $T^2$. In contrast, the same plot for SmTiO$_3$-based quantum wells shows significant non-linearity, in particular if $t_{QW} \leq 4$ SrO layers. This signals the presence of a non-Fermi liquid (NFL) regime ($n < 2$). $n$ as a function of $t_{QW}$ is shown in Fig. 2(d).

The NFL behavior can be confirmed by taking the derivative of $R_{xx}$:

$$\frac{dR_{xx}}{dT} = nAT^{n-1}. \tag{S2}$$

The derivative will give a straight line for a log-log plot of $dR_{xx}/dT$. Figure S2(c) illustrates that such a plot is indeed linear above 100 K. In the case of quantum wells with NFL behavior the slope matches the exponent extracted from fitting $\cot(\theta_H)$ and $R_H$. The downturn in $dR_{xx}/dT$ below 100 K is the manifestation of the resistivity upturn seen at low $T$. Plotting $R_{xx}$ against $T^n$ (instead of $T^2$) gives a linear slope.

Figure S3 completes Fig. 2 in the main text, by showing the results for the remaining parameters obtained from the fitting of $\cot(\theta_H)$ and $R_H$. The parameter $\alpha$ is the slope of the Hall angle, $\cot(\theta_H) = (C + \alpha T^2)$. $\alpha$ is an energy scale for the Hall scattering rate ($\tau_H$). For the quantum wells, $\alpha$ increases at small $t_{QW}$, as it approaches the metal-insulator transition. We have previously interpreted this as indication of mass enhancement [1]. The increase of $\alpha$ with decreasing $t_{QW}$ is monotonous, with no significant anomaly near 5 SrO layers (unlike for $C$ and $n$).

Figure S3 also shows the temperature slope of $R_{xx}$. Interpreting $A$ is not trivial, in the case when $n$ in $R_{xx} = R_0 + AT^n$ changes. $n$ enters into the units of $A$ and $A^{1/n}$ ($\Omega/K^n$ and $\Omega^{1/n}/K$, respectively), which thus appear divergent near 5 SrO layers owing to the discontinuity in $n$. $A$ can be converted into a temperature scale $T_0$ (independent of $n$) by re-writing the Fermi liquid formula as:

$$R_{xx} = R_0\left(1 + \left(\frac{T}{T_0}\right)^n\right), \qquad (S3)$$

where

$$T_0 = \left(\frac{R_0}{A}\right)^{1/n}. \qquad (S4)$$

As can be seen from Fig. S3, the resulting quantity $T_0$ shows no discontinuity. This leads us to conclude that the temperature slope of $\tau_{TR}$ is not a divergent quantity near 5 SrO layers.

**Supplementary figures**

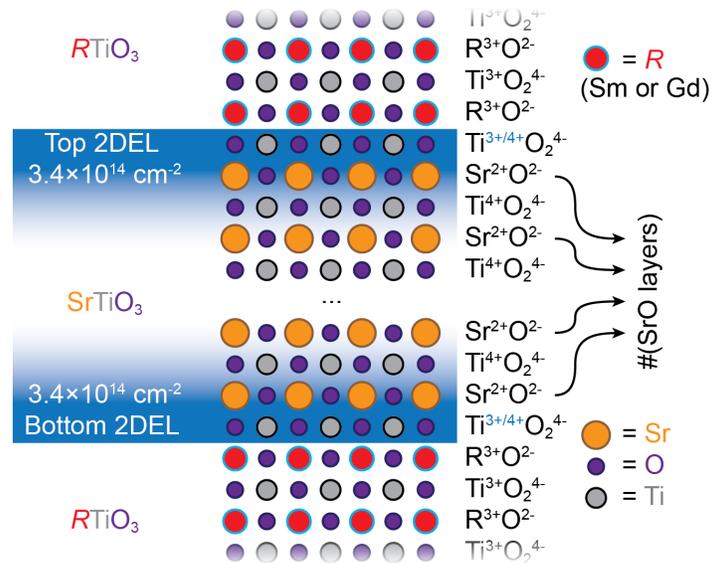

**Figure S1:** Schematic of a $R$TiO$_3$/SrTiO$_3$/$R$TiO$_3$ quantum well and 2DEL formation at the interfaces.

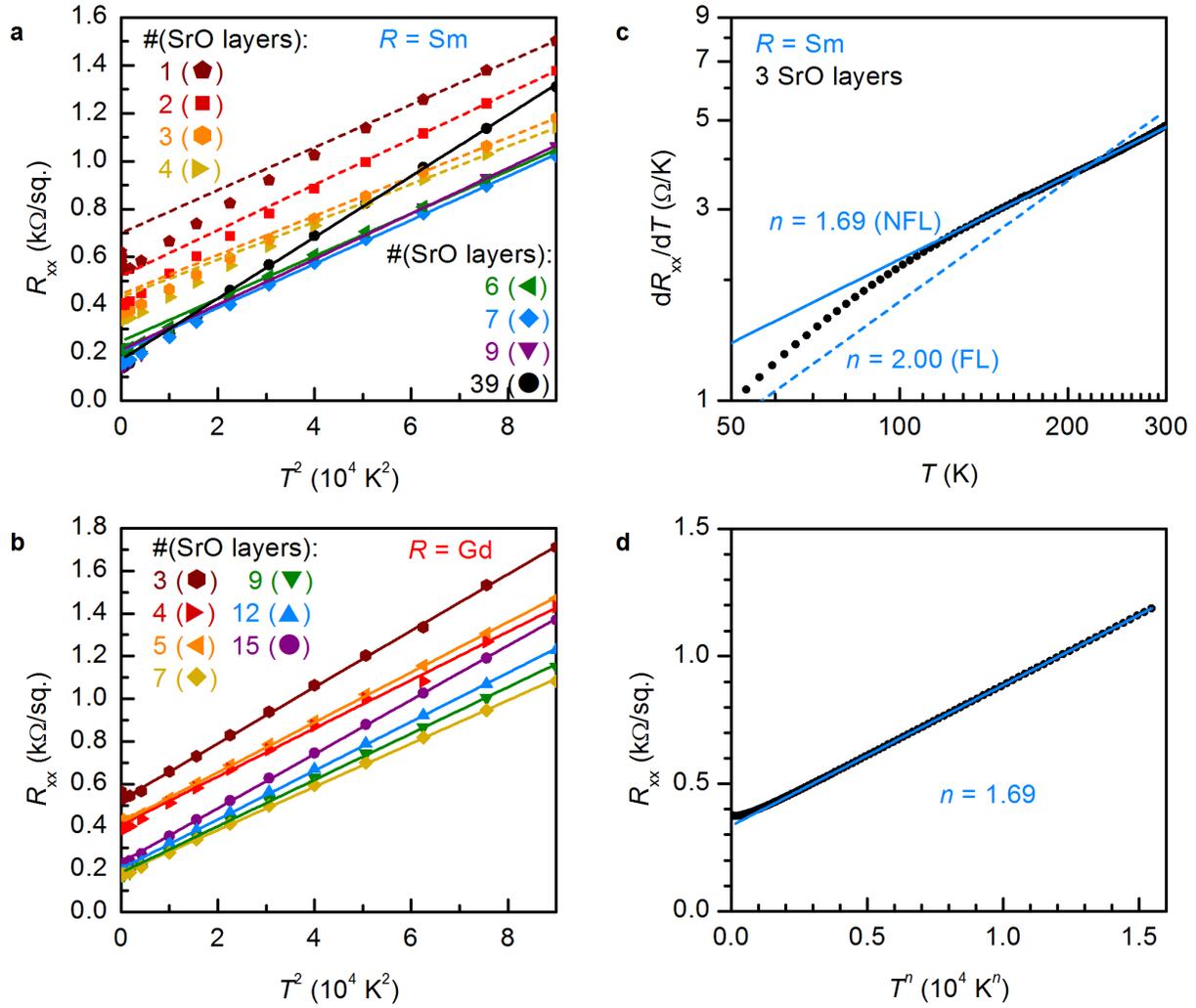

**Figure S2:** Longitudinal resistance vs. $T^2$ for (a) $R$ = Sm and (b) $R$ = Gd. (c) Temperature derivative of $R_{xx}$ for $R$ = Sm and 3 SrO layers, plotted vs. $T$ on a log-log scale. The solid line corresponds to $n$ = 1.69. The dashed line, $n$ = 2, does not fit the experimental data. (d) $R_{xx}$ for the same sample plotted vs. $T^{1.69}$.

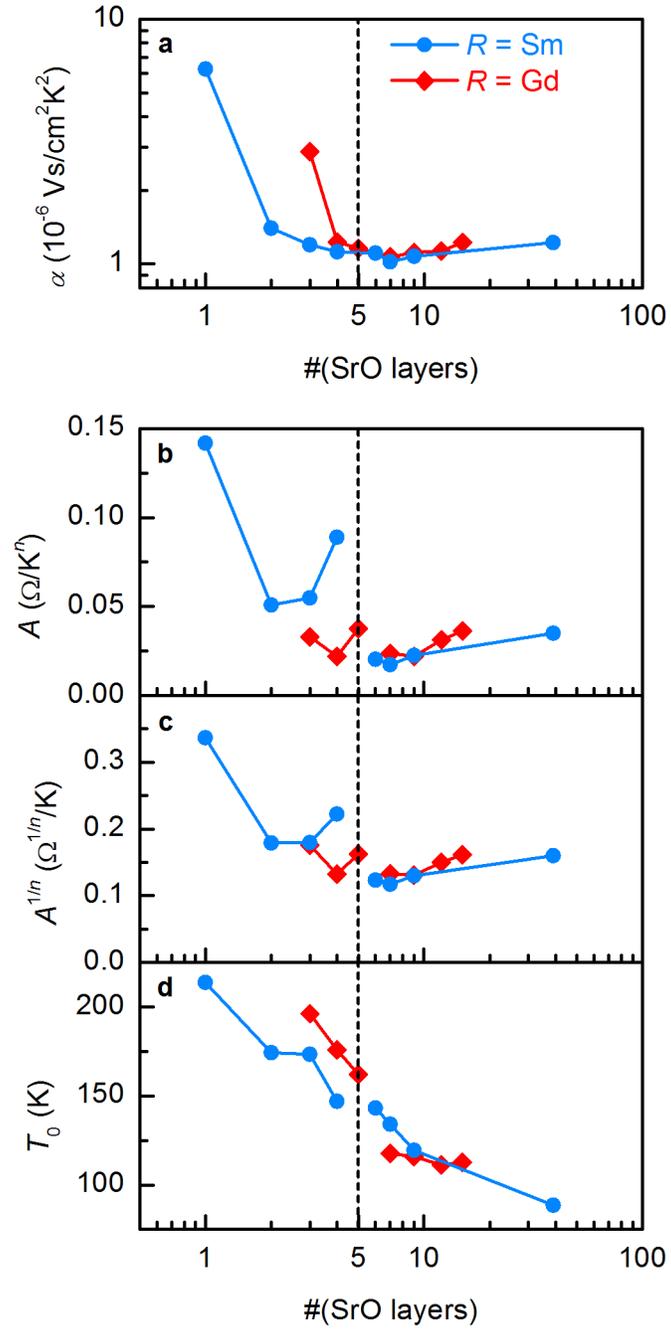

**Figure S3:** Results for the fit parameters: (a) $\alpha$, (b) $A$, (c) $A^{1/n}$ and (d) $T_0$.

# References

[1] P. Moetakef, C. A. Jackson, J. Hwang, L. Balents, S. J. Allen, and S. Stemmer, Phys. Rev. B **86**, 201102 (2012).